\documentclass[iop,twocolumn]{emulateapj}

\usepackage{natbib, aas_macros}
\usepackage{epstopdf}

\usepackage{color,ulem,soul}
\sethlcolor{hlcolor}

\renewcommand\emph[1]{\textit{#1}}
\makeatletter
\DeclareRobustCommand{\em}{%
  \@nomath\em \if b\expandafter\@car\f@series\@nil
  \normalfont \else \itshape \fi}
\makeatother

\shorttitle{Moon Migration in Weakly Accreting CPDs}
\shortauthors{Fujii, Kobayashi, Takahashi \& Gressel}

\begin{document}

\title{Orbital Evolution of Moons in Weakly Accreting Circumplanetary Disks}

\author{Yuri I. Fujii\altaffilmark{1},
    Hiroshi Kobayashi\altaffilmark{2},
    Sanemichi Z. Takahashi\altaffilmark{3},
    Oliver Gressel\altaffilmark{1}
}

\affil{
$^1$ Niels Bohr International Academy, The Niels Bohr Institute,
                Blegdamsvej 17, DK-2100, Copenhagen \O, Denmark \\
$^2$ Department of Physics, Nagoya University,
                Furo-cho, Showa-ku, Nagoya, Aichi, 464-8602\\
$^3$ Astronomical Institute, Tohoku University,
                6-3 Aramaki, Aoba-ku, Sendai, Japan, 980-8578
}

\email{yuri.fujii@nbi.ku.dk (YIF)}
\begin{abstract}
{We investigate the formation of hot and massive circumplanetary disks
  (CPDs) and the orbital evolution of satellites formed in these
  disks.  Because of the comparatively small size-scale of the
  sub-disk, quick magnetic diffusion prevents the magnetorotational
  instability (MRI) from being well-developed at ionization levels
  that would allow MRI in the parent protoplanetary disk. In the
  absence of significant angular momentum transport, continuous mass
  supply from the parental protoplanetary disk leads to the formation
  of a massive CPD. We have developed an evolutionary model for this
  scenario and have estimated the orbital evolution of satellites
  within the disk.  We find, in a certain temperature range, that
  inward migration of a satellite can be stopped by a change in the
  structure due to the opacity transitions.  Moreover, by capturing
  second and third migrating satellites in mean motion resonances, a
  compact system in Laplace resonance can be formed in our disk
  models.  }
\end{abstract}

\keywords{planets and satellites: formation --
planets and satellites: gaseous planets -- protoplanetary disks}

\section{Introduction} \label{sec:1}

Regular satellites around gas-giant planets are thought to form in a
surrounding gaseous disk. This notion is supported by the
near-circular orbits of the moon systems in our own Solar System,
which are well-aligned to their equatorial-planes, except for
irregular satellites that have been captured dynamically.

In analogy with a minimum mass solar nebula \citep{hay81} for planet
formation, minimum mass sub-nebula models were introduced for
satellite formation \citep[e.g.][]{lun82}. For explaining
certain characteristics of the Galilean moons, \citet{can02, can06} introduced
the so-called \emph{gas-starved} disk model and reproduced the ratio between the
planet mass and the total mass of the satellite system.
Satellite formation around gas
giants in a \emph{solids-enhanced} minimum mass model was discussed by
\citet{mos03a, mos03b} and \citet{est06} moreover developed a scenario
in \emph{gas-poor} environment.
\citet{sas10} introduced an inner cavity in a gas-starved disk
and found that several moons in a resonance can be formed.
Based on this work, \citet{ogi12} performed N-body
simulations and have shown that moons are commonly captured in a
2:1 mean motion resonance outside the cavity and Galilean-like
configuration can be formed.

In contrast to the \emph{nebula} hypothesis, relatively small rocky
satellites present today can be well explained by a formation scenario
in a tidal spreading disk \citep{cha10, cri12, hyo15}.  The model is
compelling, but it requires a disk of solid material as the starting
point. Such circumplanetary \emph{``debris''} disks may originate from
the capture of planetesimals \citep[e.g.][]{hyo16,hyo17} or from
tidal disruption of a previous generation of satellites.  However,
larger satellites, especially those that maintain an atmosphere, need
gas around them during their accretion.  Therefore, it is reasonable
to assume that at least some of the regular satellite systems must
have originated from gaseous circumplanetary disks (CPDs).

When a protoplanet grows to the size of several earth masses in a
protoplanetary disk (PPD) comparable to the one typically assumed for
the early solar nebula \citep{hay81}, gas around the planet starts to
accrete onto it.  At that time, because of the conservation of angular
momentum, a rotationally supported disk forms around the planet.  On
theoretical grounds, CPDs can be observed in many hydrodynamic and
magnetohydrodynamic (MHD) simulations \citep[e.g.][]{tan02, kla06,
  ayl09a, mac10, szu14, szu17, gre13, pre15}.  Because of the orders
of magnitude difference in spatial scales, however, resolving the very
vicinity of the planet in those simulations is still difficult.
\citet{tan12} have successfully measured the mass infall rate onto a
CPD during the early stage of its evolution, but the long-term
evolution remains to be established.  Yet, the modeling a CPD fully
self-consistently during the full PPD and planetary gap evolution is
difficult just like modeling the formation of PPDs from cloud-collapse
is difficult.

In PPDs, the magnetorotational instability (MRI) is thought to play an
important role in facilitating the accretion of the disk gas. Although
angular momentum transfer in a CPD was previously expected to be as
effective as that in a PPD, sustaining the MRI is more difficult in
CPDs \citep{fuj11, fuj14, tur14, kei14}.  Thermal ionization can
trigger MRI at the inner radii of a CPD if the temperature becomes
sufficiently high \citep{kei14}.  However, in the absence of strong MRI
turbulence, gas accretion may not be efficient enough to prevent a CPD
from becoming massive by accumulation of infalling material. An
alternative scenario for angular-momentum transport within the CPD may
be provided by a magnetocentrifugal disk wind which has been found to
operate sporadically in resistive-MHD simulations \citep{gre13}. It
remains to be shown whether CPD winds are equally emerging when
including additional micro-physics such as ambipolar diffusion. In any
case, disk outflows are generally competing with infall, and it is
unclear how a steady state can be reached for CPDs that are still
deeply embedded in their parent disks.

If the sub-disk grows so massive as to become gravitationally
unstable, spiral arms appear and transfer angular momentum. Whether or
not the gravitational energy is converted into heat \emph{in situ} is
still under debate, but supposed the energy deposition is local, the
temperature with the CPD can become high.  In such a situation,
episodic accretion caused by a combination of the gravitational
instability (GI) and the MRI
(boosted by thermal ionization) is to be expected.  \citet{mar11a,
  mar13} and \citet{lub12,lub13} studied this phenomenon in a layered
CPD model that is developed in the context of PPDs
\cite[e.g.][]{arm01}.

As mentioned earlier, a CPD is likely to become massive in the absence
of significant transport of angular momentum, that is if the
temperature is insufficient to maintain MRI turbulence.  In this
paper, we develop an alternative model of the satellite-forming region
of CPDs considering the mass inflow from the PPD as the dominant
factor.  Because there still is a gap between the spatial scales that
can be well-resolved by full-blown 3D MHD simulations and the actual
satellite-forming region (inside a few tens of planet radii), we pursue
the strategy of 1D modeling of the sub-disk by means of an effective
description based on results from numerical simulations.

By doing so, in some of our models, we find a bump in the radial
surface density structure.  We will examine whether such a specific
location can stop the migration of moons.  Given the situation that
the innermost moon survives rapid inward type-I migration by
convergent migration to the pressure maximum, we investigate the
possibility of trapping the second and third moons in a mean motion
resonance (MMR).  The inner three of the Galilean moons are known to be in
a 4:2:1 mean motion resonance, a so-called Laplace resonance.  In some
of our models, we successfully obtained a system in Laplace resonance.

This paper is organized as follows.  In Section \ref{sec:CPDmodel}, we
describe our sub-disk model and assumptions and resulting disks are
shown in \ref{sec:disk}. We then highlight several
models with interesting structure and discuss the orbital evolution of
moons in the disks in Section \ref{sec:migration}.  Discussion of the
obtained results and a brief summary are given in Sections
\ref{sec:discussion} \& \ref{sec:summary}, respectively.

\section{Modeling of Circumplanetary Disks} \label{sec:CPDmodel}

\subsection{Derivation of Surface Density and Temperature}
\label{sec:S-Tmodel}

The equation for the time evolution of surface density is essentially
derived in the same way as in \citet{fuj14}, but in addition, here we
simultaneously solve for the \emph{temperature} structure of the
embedded sub-disk.

We determine the surface-density profile of the CPD by solving a
diffusion equation with an additional source term stemming from mass
infall from the parent PPD. When the sub-disk's angular velocity is
taken to be Keplerian, the evolution of the surface density is given
by
\begin{eqnarray}
    \frac{\partial \Sigma}{\partial t}
    =  \frac{1}{r}\frac{\partial }{\partial r}
    \left[ 3r^{1/2}\frac{\partial}{\partial r}
            \left( r^{1/2}\nu\Sigma \right)\right]
            + f\,,
    \label{dSdt}
\end{eqnarray}
where $r$ is the radius, $\nu$ is the kinematic viscosity coefficient,
and $f$ is the mass flux from infall onto the CPD. We employ the
standard $\alpha$ prescription of \citet{sha73}, namely, $\nu = \alpha
c_{\rm s}^2/\Omega_{\rm K}$ with $c_{\rm s}$ being the sound speed and
$\Omega_{\rm K}$ the Keplerian rotation frequency.

To determine a prescription for the source term $f$, we adopted the results of
a detailed analysis of the 3D high-resolution simulations by
\citet{tan12}.  Even though they employed 11 levels of nested grids,
the resolution is insufficient to resolve the vertical structure of
the CPD in the innermost several Jupiter radii from the planet.
Therefore, \citeauthor{tan12} measured physical values of infalling
material at high altitude, where the infall is supersonic -- the idea
being that the infall rate obtained in this way is not affected by the
uncertainty caused by the architecture of the CPD further downstream.
The effective mass flux onto an inner part of CPD is $f(r) \propto
r^{-1}$ \citep[see Figure 15 of][]{tan12}.  We assume that the planet
has 0.4$\,M_{\rm J}$ and lies $5.2\,$AU away from a solar mass star.
Based on the values at this distance in the minimum mass solar nebula
\citep{hay81}, the local surface density and sound velocity of the PPD
are $\varepsilon\times 143\,{\rm g\,cm}^{-2}$ and $4.58\times10^4{\rm
  cm\, s}^{-1}$, respectively, where $\varepsilon$ is a scaling factor
representing the reduction of the surface density due to gas
dissipation.  With these values, we obtain the mass infall rate as
\begin{eqnarray}
    f = \left\{ \begin{array}{ll}
        1.3\!\times\!10^{-3}
        \varepsilon
        \left( \frac{r}{R_{\rm J}} \right)^{\!-1} {\rm g\,cm^{-2}\,s^{-1}}
           & \quad (r \leq 20 R_{\rm J})\\
         0 & \quad (r > 20 R_{\rm J}) \,,
    \end{array}\right.
    \label{infall}
\end{eqnarray}
where $R_{\rm J}$ is one Jupiter radius.  As the power-law index of
the mass infall rate drops outside $\sim\,0.04$ Hill radii from the
planet \citep{tan12}, we set $f=0$ at $r > 20 R_{\rm J}$.  The initial
value $\varepsilon=1$ corresponds to the beginning of the mass infall
and a smaller value indicates a smaller mass infall rate.  Since the
viscous timescale of the CPD is sufficiently smaller than that of the
PPD, we treat $\varepsilon$ as a constant here.

\begin{table*}[t]
    \centering
    \begin{tabular}{cccccc}
        \hline
        \hline
        opasity regime & $\kappa_0$ & $a$ & $b$ &
        \multicolumn{2}{c}{Temperature range}\\
          & (cm$^2$ g$^{-1}$) &   &  & from (K)  & to (K)\\
        \hline
        Ices & $2\times10^{-4}$ & 0 & 2 &
        $0$ & $166.8$\\
        Sublimation of ices & $2\times10^{16}$ & 0 & -7 &
        $166.8$ & $202.6$\\
        Dust & $1\times10^{-1}$ & 0 & 1/2 &
        $202.6$ & $2286.7\rho^{2/49}$\\
        Sublimation of dust & $2\times10^{81}$ & 1 & -24 &
        $2286.7\rho^{2/49}$ & $2029.7\rho^{1/81}$\\
        Molecules & $1\times10^{-8}$ & 2/3 & 3 &
        $2029.7\rho^{1/81}$ & $10000.0\rho^{1/21}$\\
        \hline
    \end{tabular}
    \caption{Bell and Lin opacity from Table 1 of \citet{kim12}}
    \label{tab:1}
\end{table*}
We adopt the following simplified form used in previous studies
\citep{can93, arm01} to estimate the temperature structure, that is
\begin{eqnarray}
    \frac{\partial T_c}{\partial t}
    = \frac{2\left( Q_+ - Q_- \right)}{c_p \Sigma}
    - v_r\frac{\partial T_c}{\partial r},
    \label{dTdt}
\end{eqnarray}
where $Q_+ = (9/8)\nu\Sigma\Omega_{\rm K}^2$ represents the viscous heating,
$Q_- = \sigma \left( 1 + 3/8\kappa \Sigma \right)T_c^4$ is the
radiative cooling, $\sigma$ is the Stefan-Boltzmann constant, and
$v_{r}$ is the radial velocity.  Further to this, the opacity $\kappa$
is given by \citet{bel94}, summarized in Table~\ref{tab:1}\citep[see
  also table~1 of][]{kim12} as
\begin{equation}
  \kappa=\kappa_0\ \rho^a\ T^b\,,
\end{equation}
where $\rho$ is the density and $T$ is the temperature.  The specific
heat is given by $c_p\simeq 2.7\,{\cal R}/\bar{\mu}$, where ${\cal R}$
is the ideal gas constant and $\bar{\mu} = 2.34$ is the mean molecular
weight.

We solve Eqns.~(\ref{dSdt}) and (\ref{dTdt}) numerically with boundary
conditions of zero torque and vanishing temperature gradient at the
inner and outer boundaries.  The temperature of the PPD at the
location of the sub-disk is assumed to be $T=123\,$K.  We set the
temperature of the CPD to this value whenever the calculated
temperature is lower than this.


\subsection{Origin of Viscosity}
\label{sec:viscosity}

In this section, we explain how to obtain an estimate for the
kinematic viscosity coefficient. The best-known origin of effective
viscosity in accretion disks is via MRI turbulence. There are two
criteria for the MRI that must be fulfilled for the instability to be
active: the disk gas must be ionized enough to be coupled with
magnetic field, and the magnetic field is not too strong \citep{bal91,
  san99, oku13a}.  We consider the MRI if the following two conditions
are satisfied:
\begin{enumerate}

\item the Elsasser number, $\Lambda \equiv v_{{\rm A}z}^2/\eta
  \Omega_{\rm K}>1$, with $v_{{\rm A}z} \equiv |B_z|/\sqrt{4\pi \rho}$
  being the vertical component of the Alfv\'en velocity, and where
  $\eta$ is the magnetic diffusivity, and

\item the $z$ component of plasma beta defined in terms of net
  magnetic flux, $\beta_{\rm z}=2c_{\rm s}^2/v_{{\rm
      A}z}^2\gtrsim2000$; see
  \citet{oku13a} and \citet{fuj14} for details.

\end{enumerate}
As suggested by the results of \citet{fuj11, fuj14}, \citet{tur14},
and \citet{kei14}, sub-disks are not likely to widely sustain
well-developed MRI turbulence in the absence of thermal ionization.

\begin{figure}[t]
  \plotone{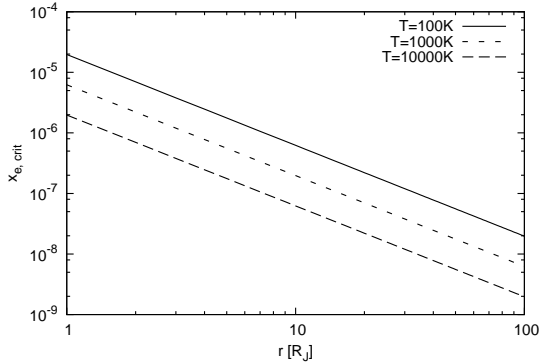}\\[-8pt]
  \caption{Critical ionization degree needed to sustain the MRI at the
    midplane at $T=100\,$K, 1000$\,$K, and 10,000$\,$K for
    $\beta_0=10^5$.}
  \label{fig:mri}
\end{figure}

If the disk is not subject to sustaining MRI turbulence, material may
pile up and the disk eventually becomes gravitationally
unstable. Accordingly, if the Toomre parameter,
\begin{equation}
  Q \equiv \frac{\Omega_{\rm K} c_{\rm s}}{\pi G\, \Sigma}\,,
\end{equation}
(with $G$ the gravitational constant) becomes smaller than a value of
2, we employ an effective viscosity of $\alpha_{\rm GI} \equiv
\exp(Q^{-4})$ \citep{zhu10,tak13}.  In such a case, the disk can be
easily heated up provided that gravitational energy is converted into
heat.  It is widely assumed that thermal ionization can trigger the
MRI if the temperature exceeds about $1000\,$K; however, in CPDs this
is not always the case. First of all, the ionization fraction obtained
from the Saha equation depends on density and a gravitationally
unstable disk naturally has a high surface density. Secondly, for a
CPD to sustain the MRI, the required ionization fraction is
comparatively higher than that of a PPD even if the density is the
same.

As is mentioned in \citet{fuj14}, this is because typical length scale
is orders of magnitude smaller in a sub-disk. Thus, the critical
temperature is higher than $1000\,$K. If we adopt $\eta = 234(T/1 {\rm
  K})^{1/2}\ x_{\rm e}^{-1}\,{\rm cm}^2{\rm s}^{-1}$ where $x_{\rm
  e}\equiv n_{\rm e}/n_{\rm n}$ is ionization degree and $n_{\rm e}$
and $n_{\rm n}$ are respectively number density of electron and neutral
gas \citep{bla94}, the condition for sufficient ionization to sustain
the MRI at midplane can be given as
\begin{eqnarray}
    \Lambda =\frac{2c_{\rm s}^2x_{\rm e}}
            {234\sqrt{T/1 {\rm K}}\,\Omega_{\rm K}\beta_0}>1,
    \label{MRI}
\end{eqnarray}
where $\beta_0$ is the plasma beta at the midplane.
From this equation, we can derive the critical ionization degree
needed to be MRI-active as
$x_{\rm e,crit}=234\sqrt{T/1 {\rm
  K}}\Omega_{\rm K}\beta_0/(2c_{\rm s}^2)$.

As an illustration of this, Fig.~\ref{fig:mri} shows the critical
ionization degree needed to have the MRI at the midplane for a disk
with $\beta_0=10^{5}$.  The plot is made for the disk temperatures of
100$\,$K, 1000$\,$K, and 10,000$\,$K, respectively. Note that,
with the ``gas-starved'' case of surface density as the lower limit,
$\Sigma=100\ (r/20R_{\rm J})^{-3/4}\ {\rm g\ cm^{-2}}$
\citep{can02,sas10},
midplane ionization degree at 10$\,R_{\rm J}$, when
$T=1000\,$K is assumed is only about $10^{-12}$.  One can easily
see from Fig.~\ref{fig:mri} that this value is far below the critical
value, $x_{\rm crit}$, at the respective radius.
The temperature needed to obtain $x_e>x_{\rm e, crit}$ for this case
is about 2000$\,$K. Obviously, if the surface density is larger,
higher temperature is required.

\begin{figure*}[t]
    \plottwo{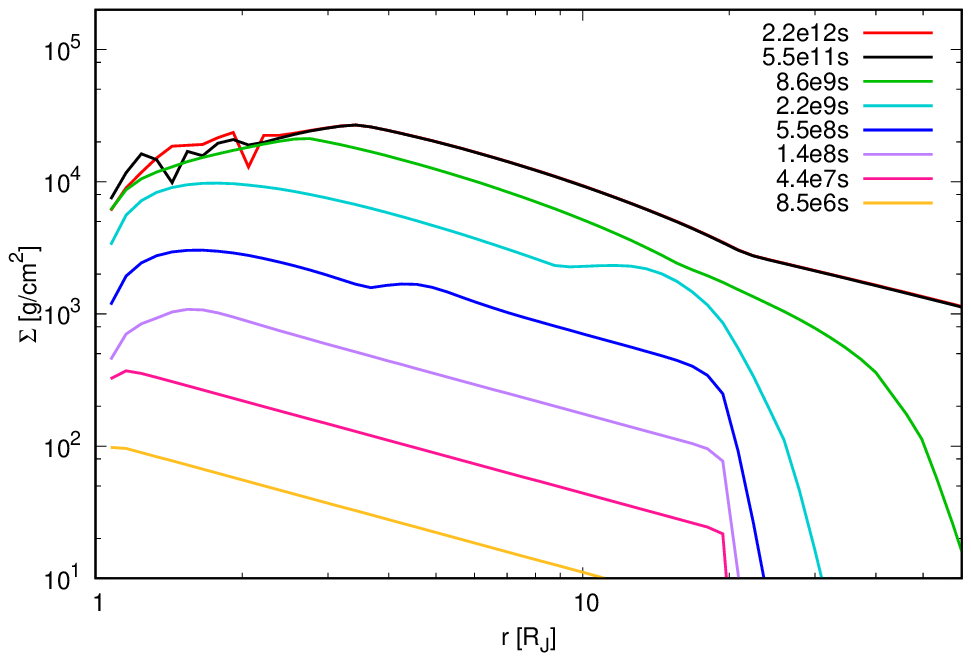}{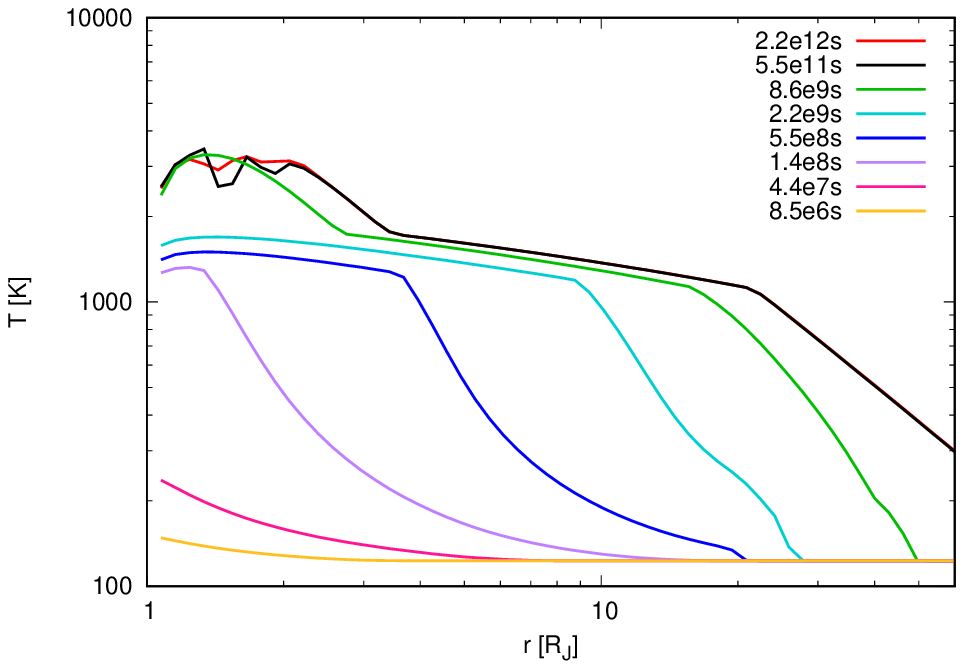}\\[-5pt]
         \caption{Time evolution of surface density (left) and temperature
         structure (right) for the case with
         $\varepsilon=10^{-2}$ and $\alpha_{\rm floor} = 10^{-4}$.
         Each color shows the different time of the evolution.
         The uppermost line in each plot shows the
         steady-state value except within 2.2 $R_{\rm J}$.}
         \label{fig:evolution}
         \medskip
\end{figure*}

In this paper, we consider the thermal ionization of potassium,
sodium, and magnesium.  In the high-temperature regime, where most of
the metals are already ionized, atomic hydrogen also becomes a
dominant source of free electrons.  At such temperatures, hydrogen gas
is already dissociated, so we employ the following Saha equation for
atomic hydrogen as well as for metals to solve for the ionization
degree from collisions\footnote{These expressions are approximately
correct when the ionization fraction of each species is small,
which is not appropriate for K and Na in this case. Since their abundance
is small, however, the resulting ionization degree is not strongly affected.}:
\begin{eqnarray}
    \frac{x_{\rm e}x_{\rm K^+} }{x_{\rm K}}&=&\frac{2}{n_{\rm n}}
                   \left(\frac{2\pi m_{\rm e} k_{\rm B}T_c}
                    {h^2}\right)^{3/2}\nonumber\\
                  &\ & \times \exp(-5.0\times10^4/T_c)\\
    \frac{x_{\rm e}x_{\rm Na^+} }{x_{\rm Na}}&=&\frac{2}{n_{\rm n}}
                   \left(\frac{2\pi m_{\rm e} k_{\rm B}T_c}
                    {h^2}\right)^{3/2}\nonumber\\
                  &\ & \times \exp(-6.0\times10^4/T_c)\\
    \frac{x_{\rm e}x_{\rm Mg^+} }{x_{\rm Mg}}&=&\frac{2}{n_{\rm n}}
                   \left(\frac{2\pi m_{\rm e} k_{\rm B}T_c}
                    {h^2}\right)^{3/2}\nonumber\\
                  &\ & \times \exp(-8.9\times10^4/T_c)\\
    \frac{x_{\rm e}x_{\rm H^+} }{x_{\rm H}}&=&\frac{2}{n_{\rm n}}
                   \left(\frac{2\pi m_{\rm e} k_{\rm B}T_c}
                    {h^2}\right)^{3/2}\nonumber\\
                  &\ & \times \exp(-1.6\times10^5/T_c)
    \label{saha}
\end{eqnarray}
where, $x_{\rm K}$ etc. represent the number densities of each
species, and $m_{\rm e}$, $h$, and $k_{\rm B}$ are respectively
electron mass, the Planck constant, and the Boltzmann constant.  With
thoes equations and charge neutrality, we finally obtain the
ionization degree as
\begin{eqnarray}
    x_{\rm e} = & \left( \frac{2}{n_{\rm n}} \right)^{\frac{1}{2}} &
                    \left(\frac{2\pi m_{\rm e} k_{\rm B}T_c}
                    {h^2}\right)^{3/4}\qquad \nonumber\\
      \times \Big[ \Big.\quad
            & x_{\rm K} & \exp\left(\frac{-5.0\times10^4}{T_c} \right)
           \nonumber\\
         + & x_{\rm Na} & \exp\left(\frac{-6.0\times10^4}{T_c} \right)
           \nonumber\\
         + & x_{\rm Mg} & \exp\left(\frac{-8.9\times10^4}{T_c} \right)
           \nonumber\\
         + & x_{\rm H} & \exp\left(\frac{-1.6\times10^5}{T_c} \right)
      \Big.\Big] ^{\frac{1}{2}} .
    \label{xe_th}
\end{eqnarray}
We adopt Solar abundance multiplied by depletion factor $\delta$ for
metal species, $x_{\rm K}=9.87\times10^{-8}\,\delta$, $x_{\rm
  Na}=1.60\times10^{-6}\,\delta$, and $x_{\rm
  Mg}=3.67\times10^{-5}\,\delta$, in our calculation.

If $\Lambda>1$ is satisfied at the midplane, we set the viscous
parameter due to the MRI turbulence as $\alpha_{\rm MRI} =
1040/\beta_0 + 0.015$ \citep{oku11}. Thus, we denotethe the Elsasser
number at the midplane as $\Lambda$ here after. What if the ionization
degree is not high enough to sustain the MRI and the surface density is
smaller than the critical value to be gravitationally unstable? There is
always molecular viscosity, but it is negligibly small.  Gravitational
interaction between the star, planet and gas of a CPD can be an origin
of angular momentum transport \citep{riv12}.  Kelvin-Helmholtz-like
instabilities between sedimenting dust layers and gas can generate
turbulence which can be roughly estimated as $\sim
10^{-4}-10^{-3}$. Moreover, it has been found that in disks with
imposed radial temperature gradients, the resulting vertical shear can
be a robust source of turbulence via an analog of the
Goldreich-Schubert-Fricke instability \citep{nel13}. Rigorously
establishing the presence of the vertical-shear instability in CPDs
will, however, require us to derive constraints on radiative cooling
timescales $\tau_{\rm crit}$ similar to the work by \citet{lin15}, who
(in the context of PPDs) find the corresponding criterion to scale
with the disk thickness -- which is favorably large for the
comparatively puffed-up CPDs, implying less-restrictive conditions on
$\tau_{\rm crit}$.

There may be other ways of transporting angular momentum, however,
those mechanisms contain uncertainties and the specific value is not
yet obtained. Thus, we treat them via setting a floor value,
$\alpha_{\rm floor}$, in this work.  In summary, we define the viscous
parameter as
\begin{eqnarray}
    \alpha &=& \alpha_{\rm MRI} + \alpha_{\rm GI} + \alpha_{\rm
      floor}\\[6pt]
    \alpha_{\rm MRI} &=& \left\{
                        \begin{array}{ll}
                            \frac{1040}{\beta_0} + 0.015 & \quad ({\rm
                              if\,}\Lambda>1) \\
                            0 & \quad ({\rm if\,}\Lambda\le 1)
                        \end{array}
                        \right.\\[6pt]
    \alpha_{\rm GI} &=& \left\{
                        \begin{array}{ll}
                            \exp(Q^{-4}) & \quad ({\rm if\,} Q<2) \\
                            0            & \quad ({\rm if\,} Q \ge 2)
                        \end{array}
                       \right.
    \label{alpha}
\end{eqnarray}
We take $\varepsilon$ and $\alpha_{\rm floor}$ as parameters and
obtain structures of CPDs based on Section \ref{sec:S-Tmodel} above.

\section{Resulting Disk Structure} \label{sec:disk}

Since the timescale for $\varepsilon$ to drop is uncertain, we simply
develop a disk for each parameter set from scratch until it reaches a
quasi-stationary state. First, we calculate assuming only 1\% of
metals are in gas phase, i.e. $\delta=0.01$.  Figure
\ref{fig:evolution} is an example of the formation of a CPD with
$\varepsilon=10^{-2}$ and $\alpha_{\rm floor} = 10^{-4}$.  Both
surface density and temperature increase with time.  The uppermost
lines in Figure \ref{fig:evolution} show the values in the steady
state except for within $\sim 2 R_{\rm J}$, where the radial profiles
of the surface density and temperature remain non-steady and wiggle
about.

\begin{figure}
    \plotone{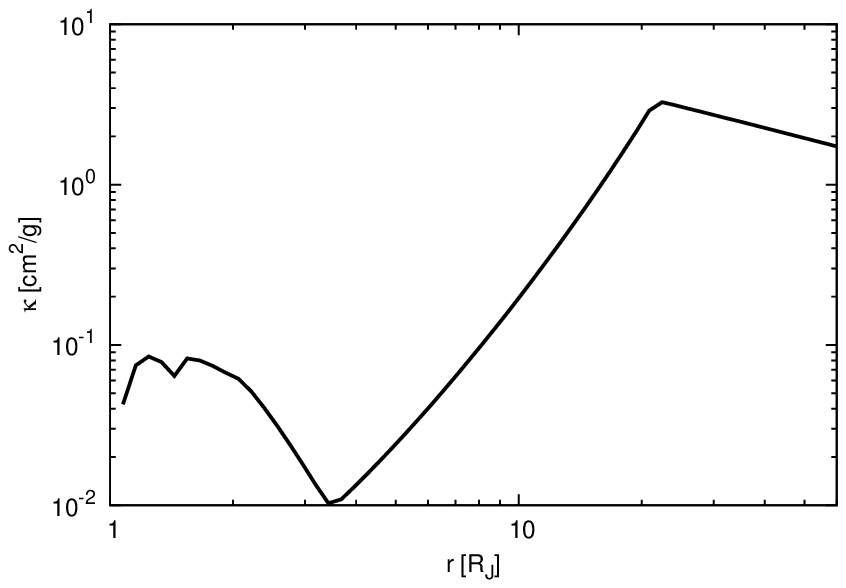}\\[-10pt]
    \plotone{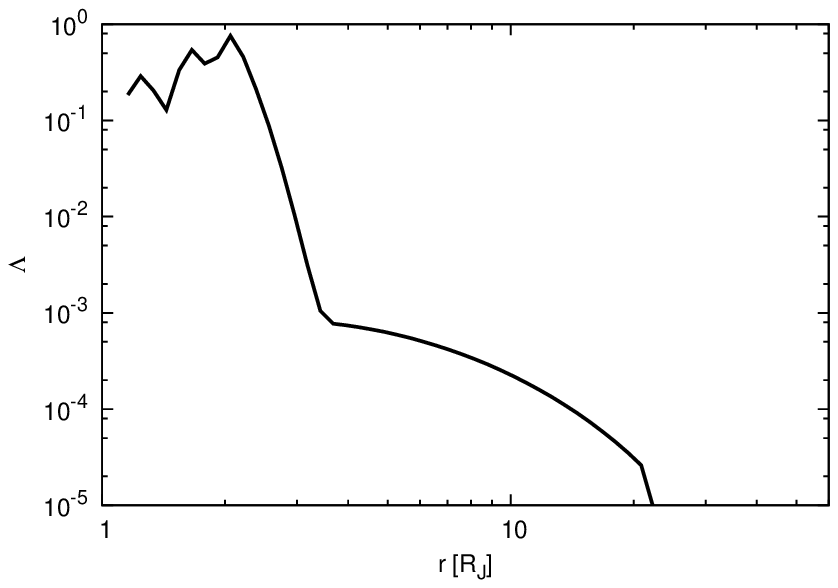}\\[-10pt]
    \plotone{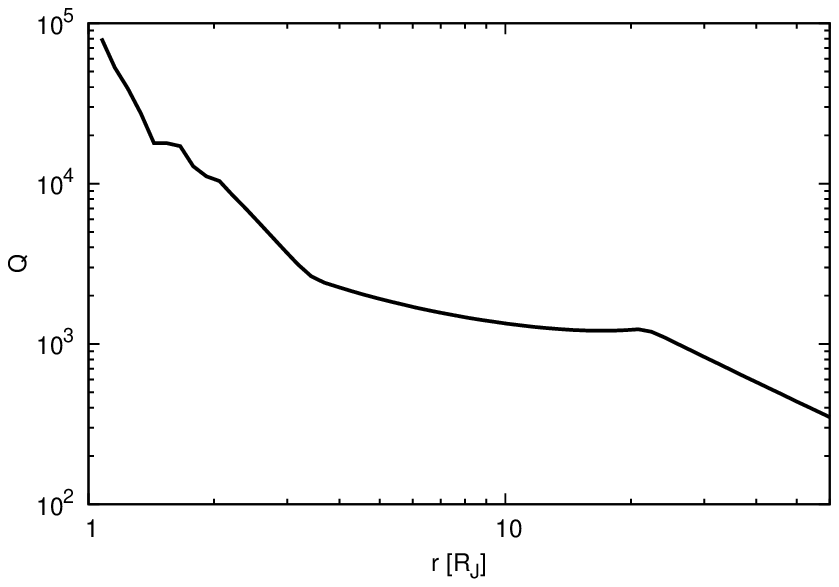}\\[-5pt]
         \caption{Radial profiles of opacity (top), Elsasser number
           (middle), and Toomre's Q parameter (bottom panel) for the
           fiducial case with $\varepsilon=10^{-2}$ and $\alpha_{\rm
             floor} = 10^{-4}$.}
         \label{fig:opacity}
\end{figure}

In Fig.~\ref{fig:opacity}, the opacity, Elsasser number, $\Lambda$,
and Q value at the final stage in Fig.~\ref{fig:evolution} are shown.
In the quiescent outer disk, the Elsasser number and Q parameter
remain in the stable regime, that is $\Lambda<1$ and $Q>2$, and
accordingly $\alpha$ is determined by $\alpha_{\rm floor}$.  The
Elsasser number occasionally exceeds unity in the inner disk and that
prevents the system from settling into a stationary solution.  Even if
the inner disk remains time-variable, the outer disk achieves a steady
state independent of the inner region.  In the quiescent outer disk, a
bump in the surface-density profile forms because of the increase of
opacity due to the transition of the origin from dust sublimation
to molecules. The dips in surface density (and
temperature) at the inner domain border are related to the boundary
conditions. Thus, we do not consider them as a bump.

%

\begin{figure*}[t]
    \epsscale{0.9}
    \plottwo{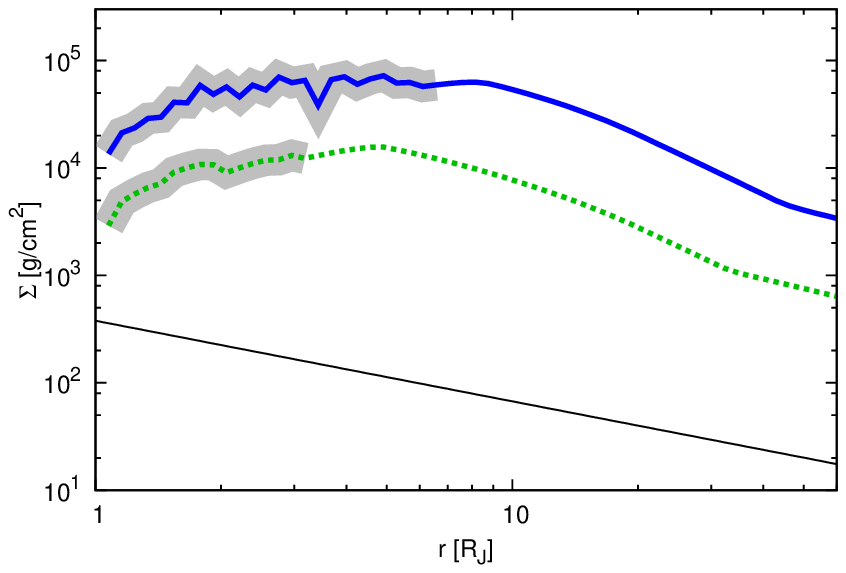}{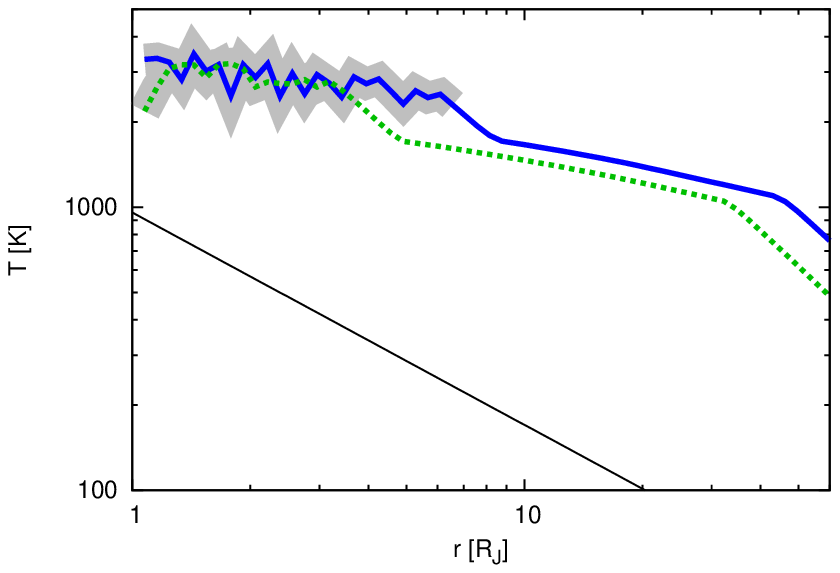}\\[-10pt]
    \plottwo{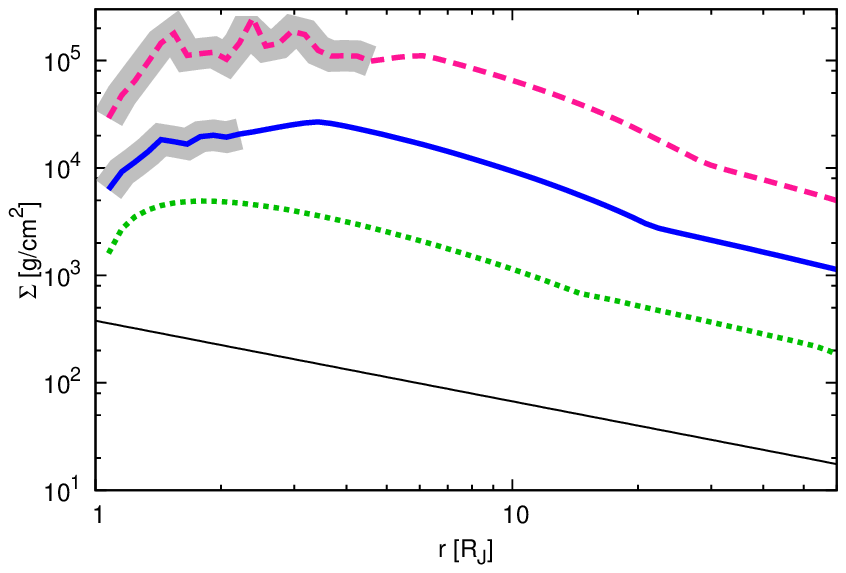}{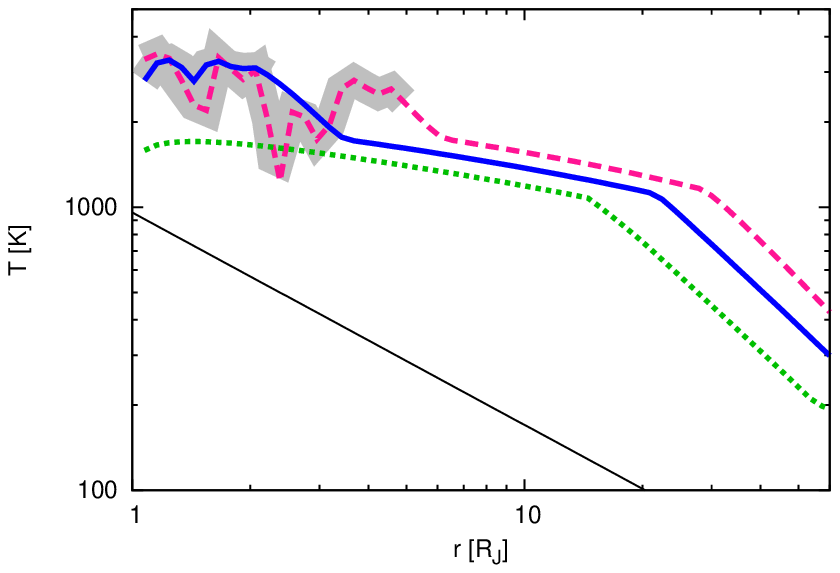}\\[-10pt]
    \plottwo{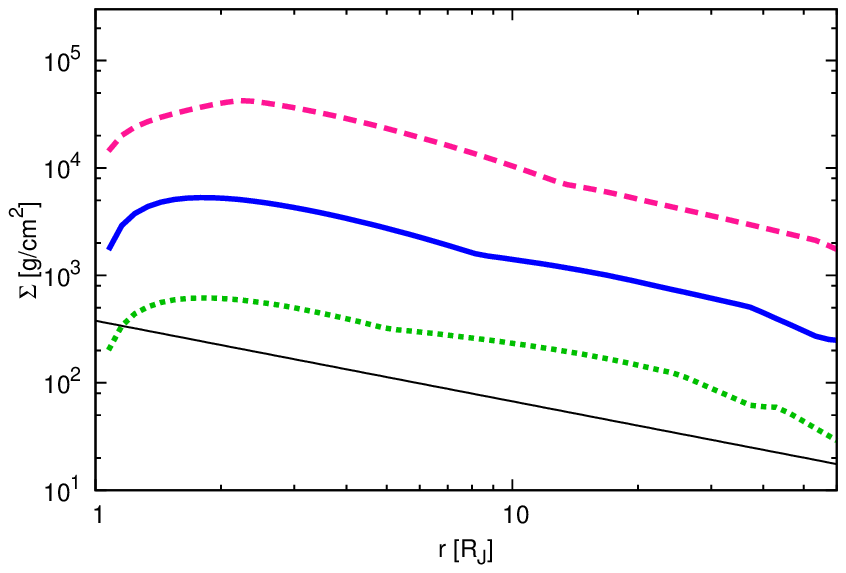}{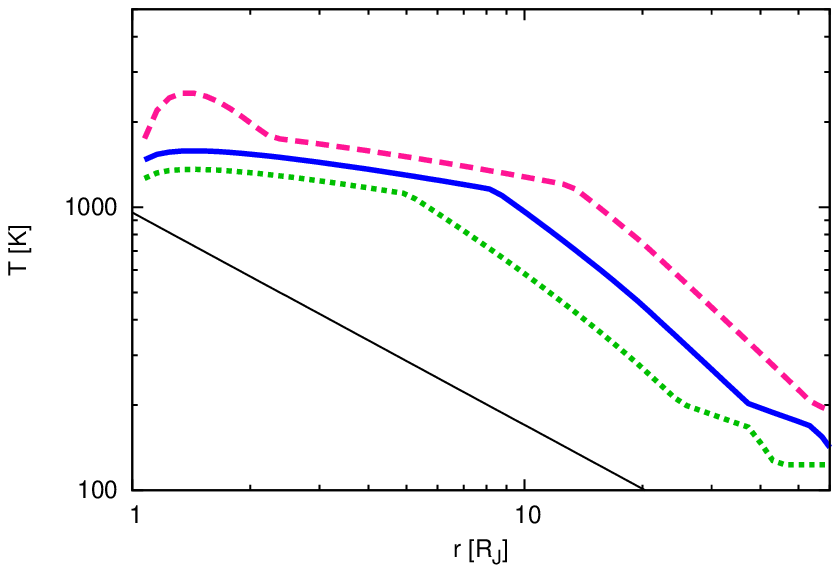}\\[-5pt]
         \caption{Steady-state surface density (left) and temperature
           (right) for $\varepsilon=10^{-1}$, $10^{-2}$, and $10^{-3}$
           (from top to bottom) with $\delta=0.01$ and various
           $\alpha_{\rm floor}$: dashed (pink), thick-solid (blue),
           dotted (green) lines show the case for $\alpha_{\rm floor}
           = 10^{-5}$, $10^{-4}$, and $10^{-3}$, respectively. Shaded
           regions are not time-steady and display jitter.  As a
           reference, a typical \emph{gas-starved} disk profile is
           also plotted (solid black line).}
         \label{fig:cpd-ep}
         \smallskip
\end{figure*}

Figure~\ref{fig:cpd-ep} shows the surface density profiles (left
column) and temperature (right column) of various disk models once the
outer disk has reached a steady state. One can see that models with
larger $\varepsilon$ and/or smaller $\alpha_{\rm floor}$ generally
become more massive and hotter.  The top panels of
Fig.~\ref{fig:cpd-ep} show the surface density and temperature
structure for $\varepsilon=10^{-1}$.  Because the ionization degree
reaches a near-critical value at the inner disk radii, the gas
accretion rate fluctuates in time (illustrated by the shaded area in
Fig.~\ref{fig:cpd-ep}), and the disk structure is not fully stationary
in this regime.  When MRI enabled by thermal ionization is developed,
depending on the settings, the disk either ends up with a steady state
with smaller surface density or enters the gravito-magneto limit cycle
studied by \citet{lub12, lub13}.

The middle panels of Fig.~\ref{fig:cpd-ep} show the disk structure for
$\varepsilon=10^{-2}$.  For the case of $\alpha_{\rm floor}=10^{-3}$
(green dotted line), relatively effective gas accretion keeps the
surface density smaller than the cases for $\alpha_{\rm
  floor}=10^{-4}$ (blue solid line) and $\alpha_{\rm floor}=10^{-5}$
(pink dashed line).  Since the temperature does not become high enough
with $\alpha_{\rm floor}=10^{-3}$, the MRI is not triggered by thermal
ionization and the whole disk settles into a steady state.  Compared
to the top panels, values are generally slightly smaller.

The bottom panel of Fig.~\ref{fig:cpd-ep} illustrates a case where the
reduction factor decreases down to $10^{-3}$.  Because the infall flux
is already small enough, the surface density does not become that
massive, and therefore the temperature cannot be as high as supplying
sufficient ionization to sustain the MRI.  We remark that the
surface-density range of our models is similar to the extended outer
disk of \citet{mos03a}, but temperature is much higher in our models.
For some parameter sets, a bump can be seen in surface density that is
formed due to the change in opacity. A radial pressure bump cannot be
seen in a disk with small $\varepsilon$ and/or large $\alpha_{\rm
  floor}$ because the surface density does not pile-up sufficiently
for the inner disk to transition into the higher temperature regime
required for the opacity transition.

\begin{figure*}[t]
    \epsscale{0.9}
\plottwo{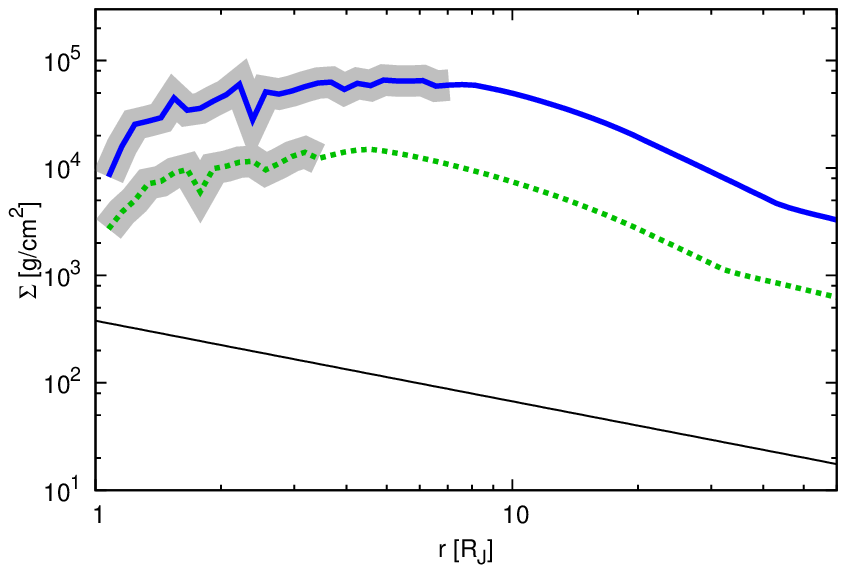}{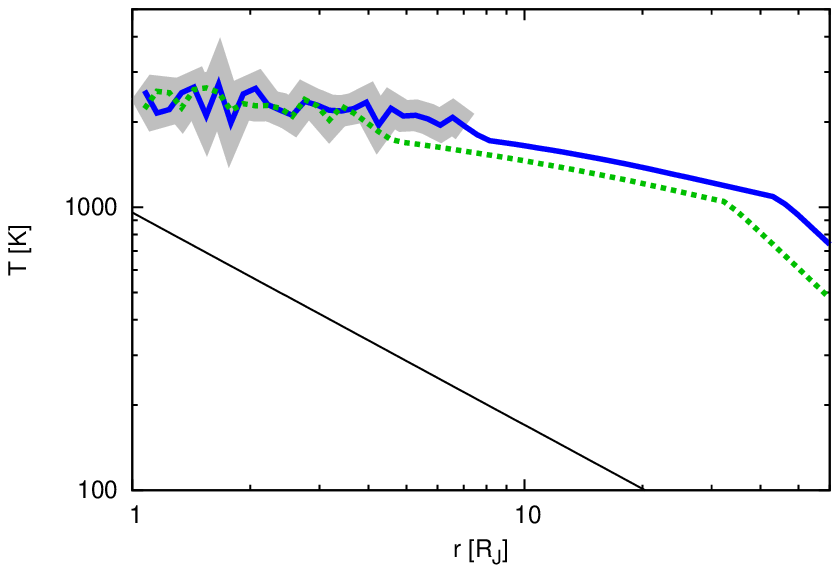}\\[-12pt]
\plottwo{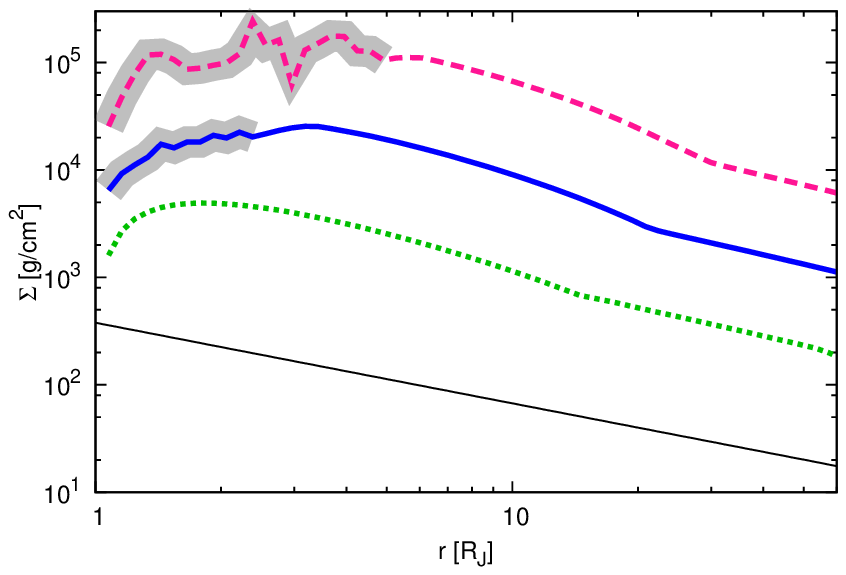}{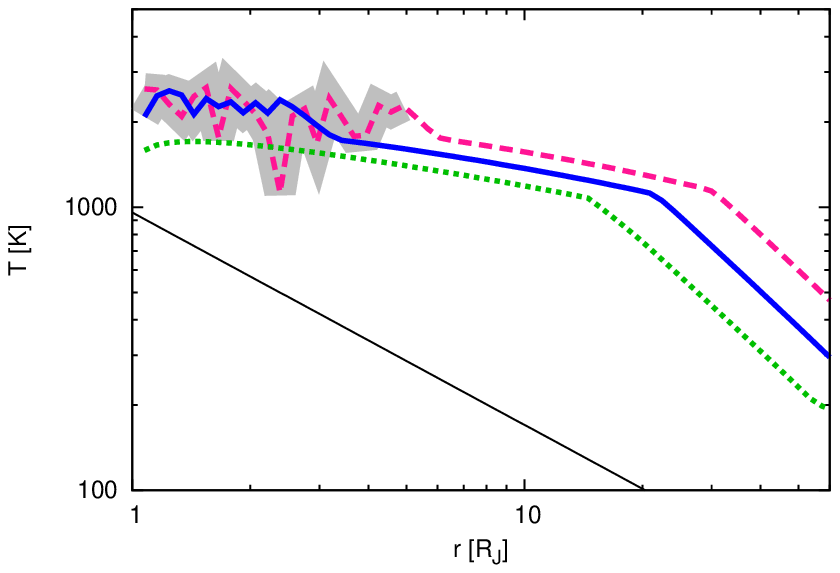}\\[-6pt]
\caption{Same as Fig.~\ref{fig:cpd-ep} but for $\delta=1$, that is,
  undepleted composition.  The top panels are for
  $\varepsilon=10^{-1}$ and the bottom panels are for
  $\varepsilon=10^{-2}$.  The results for $\varepsilon=10^{-3}$ are
  not shown because they are identical to Figure \ref{fig:cpd-ep}.}
\label{fig:cpd-ep-dep1}
\medskip
\end{figure*}

Next, for the case when the temperature at which the thermal
ionization plays a role for gas accretion is as high as the one for
grains to evaporate, we calculate disk structures with $\delta=1$
corresponding to solar abundance.  We show the respective results in
Fig.~\ref{fig:cpd-ep-dep1}.  For models with $\varepsilon=10^{-3}$
(not show in the figure), we obtained identical profiles as in the
case with $\delta=0.01$ for all values of $\alpha_{\rm floor}$.  This
is also true for the model with $\varepsilon=10^{-2}$ and $\alpha_{\rm
  floor}=10^{-3}$.  For all other models, the range in which we do not
obtain steady-state solutions slightly increase because thermal
ionization can provide more electrons at the same temperature.

We conclude that the transition of the opacity regime from sublimation
of dust to molecules can produce interesting structures in embedded
CPDs for a variety of reasonable disk models.
In the following section, we discuss the orbital
evolution of (proto-)satellites in the disk models derived here.

\section{Capture of Satellites in Resonant Orbits} \label{sec:migration}

In the context of protoplanetary disks, disk-planet interaction has
been studied extensively in the literature. Planets are believed to
migrate in the hosting disk by exchanging their angular momentum with
the disk.  The idea is also introduced in satellite formation
\citep[for instance in][]{can02, can06, sas10, ogi12}.  As for PPDs,
the migration direction and speed depend on the satellite mass and
the disk structure, and the timescale is given as
\begin{equation}
    t_{\rm m} = \frac{1}{|b|}\left(\frac{M_s}{M_{p}}\right)^{-1}
    \left(\frac{\Sigma r^2}{M_{p}}\right)^{-1}
    \left(\frac{c_{s}}{v_{K}}\right)^{2}\Omega_{\rm K}^{-1},
    \label{t_a}
\end{equation}
where $M_{\rm s}$ and $M_{\rm p}$ are the mass of the satellite and
planet, respectively, and $b$ is a constant that determines the
direction and speed of the migration \citep{paa11,kre12,ogi15}.
In this work, we only consider moons with circular orbits, and
we use the formula of \citet{paa11}\footnote{In their formula, 
  which is derived
  as a fit to a set of simulations with a single perturbing body, the
  resulting torque from the entire disk is considered. For systems of
  multiple embedded bodies, the formula remains valid as long as all
  masses remain low enough, such that non-linear wake interaction can
  be ignored.} for the migration constant $b$ 
\citep[see also][]{ogi15}:
\begin{eqnarray}
    b &=& \frac{2}{\gamma}\left\{-2.5-1.7q+0.1p\right.\nonumber\\
          & & +1.1F(P_\nu)G(P_\nu)\left(\frac{3}{2}-p\right)\nonumber\\
          & & +0.7\left(1-K(P_\nu)\right)\left(\frac{3}{2}-p\right)\nonumber\\
          & & +7.9\frac{q-(\gamma-1)p}{\gamma}
                  F(P_\nu)F(P_\chi)\sqrt{G(P_\nu)G(P_\chi)}\nonumber\\
          & & +\left(2.2-\frac{1.4}{\gamma}\right)
                    \left[q-(\gamma-1)p\right]\nonumber\\
          & &     \left. \times\sqrt{(1-K(P_\nu))(1-K(P_\chi))} \right\},
    \label{b}
\end{eqnarray}
where $\gamma$ is the adiabatic constant,
$p \equiv -{\rm d\,ln}\ \Sigma/{\rm d\,ln}\ r$, and
$q \equiv -{\rm d\,ln}\ T/{\rm d\,ln}\ r$. The expressions for
$F(P)$, $G(P)$, and $K(P)$ are furthermore given as
\begin{eqnarray}
    F(P) &=& \Big(1+\left( \frac{P}{1.3}\right)^{\!2}\Big)^{-1} \\
    \label{F}
    G(P) &=& \left\{
            \begin{array}{ll}
                \frac{16}{25}\left( \frac{45\pi}{8}\right)^{3/4}
                P^{3/2}  & \quad P<\sqrt{\frac{8}{45\pi}} \\
                1-\frac{9}{25}\left( \frac{8}{45\pi}\right)^{4/3}
                P^{-8/3}  & \quad P\geq\sqrt{\frac{8}{45\pi}}
            \end{array}
            \right.\\
    K(P) &=& \left\{
            \begin{array}{ll}
                \frac{16}{25}\left( \frac{45\pi}{28}\right)^{3/4}
                P^{3/2}  & \quad P<\sqrt{\frac{28}{45\pi}} \\
                1-\frac{9}{25}\left( \frac{28}{45\pi}\right)^{4/3}
                P^{-8/3}  & \quad P\geq\sqrt{\frac{28}{45\pi}}\,.
            \end{array}
            \right.
\end{eqnarray}
We assume $\gamma=1.4$ and that the thermal diffusivity is the same as
$\nu$ (i.e., Pr=1).  Thus, with the dimensionless half-width of the
horseshoe region, $\chi_{\rm s} = 1.1/\gamma^{1/4}\sqrt{M_s r/M_p h}$
(where $h$ is the scale-height of the disk),
\begin{equation}
    P_\nu = \frac{2}{3}
    \sqrt{\frac{\Omega_{\rm K} r^2\chi_{\rm s}^3}{2\pi\nu}}
    = \frac{2}{3} P_\chi.
    \label{P_nu}
\end{equation}
%
\begin{table}
    \centering
    \begin{tabular}{lcc}
      \hline
         & $\varepsilon$ & $\alpha_{\rm floor}$\\[2pt]
      \hline\hline
      \\[-6pt]
        Models 1 \& 1' & $10^{-1}$ & $10^{-4}$\\
        Models 2 \& 2' & $10^{-1}$ & $10^{-3}$\\
        Models 3 \& 3' & $10^{-2}$ & $10^{-5}$\\
        Models 4 \& 4' & $10^{-2}$ & $10^{-4}$\\
        Models 5 \& 5' & $10^{-3}$ & $10^{-5}$\\
        \hline
    \end{tabular}
    \caption{Model parameters for the cases considered for orbital migration}
    \label{tab:models}
\end{table}
\begin{figure}[h]
    \plotone{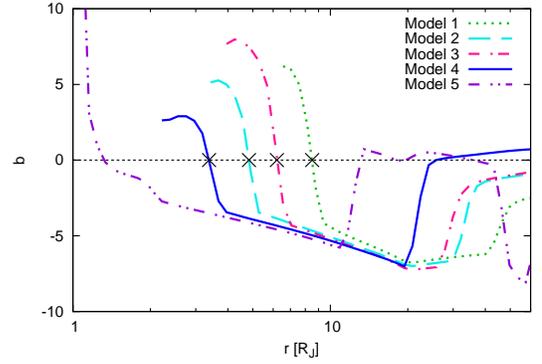}\\[-5pt]
    \caption{Migration coefficient for an Io-sized moon in the disk
      model given in Section~\ref{sec:CPDmodel} for Models 1-5. The
      positions where the migration stops for each model are indicated
      by crosses.}
    \label{fig:b}
\end{figure}
\begin{figure}[h]
    \plotone{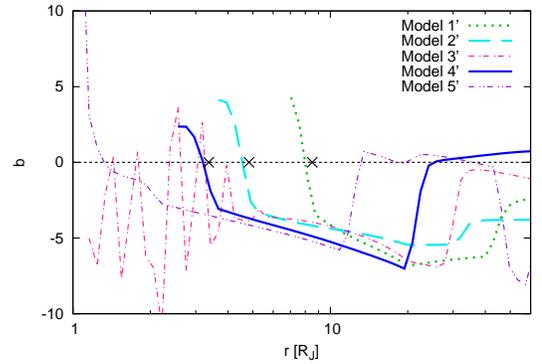}\\[-5pt]
    \caption{Migration coefficient for an Io-sized moon
    for Models~1'-5'. For the comparison, the positions
    where the first moons stop migration for Models 1,
    2, and 4 are marked with cross signs.}
    \label{fig:b-dep1}
\end{figure}
If $b$ is negative, the satellite migrates toward the planet and
positive $b$ means outward migration.  We selected disk models with a
discernible bump in surface density structure, as summarized in
Table~\ref{tab:models}. In the following, we refer to the cases with
$\delta=0.01$ as Models 1-5, and $\delta=1$ as Models 1'-5'.  The
radial distribution of $b$ for Models 1-5 with a satellite of Io mass
is given in Figure~\ref{fig:b}.  If a Io-sized satellite that formed
at outer radii migrates inward, the migration halts at locations
marked with crosses in Fig.~\ref{fig:b}.
Moons migrate all the way to
the planet in Model 5 because there is no location where the value of
$b$ changes from positive to negative as $r$ increases.
\smallskip

Figure~\ref{fig:b-dep1} shows plots of $b$ for Models 1'-5' with
$\delta=1$, that is, solar abundance.  The location for the
termination of moon migration is slightly further inside in Models 1',
2', and 4' compared to Models 1, 2, and 4, respectively, whereas the
convergence region disappears in Model 3' as compared to Model 3.  If
a satellite migrates and stays at $b\simeq0$, the second satellite
migrating from the outer disk approaches the first one that is trapped
inside the location of convergent migration.  If the migration
timescale of the second satellite is longer than the critical time
scale, $t_{\rm m}^{\rm crit}$, at the location of the first satellite, the
second satellite is captured in a mean motion resonance.
As mentioned in \citet{ogi13}, the capture probability for higher-order
resonances is very low, and moreover the 2:1 MMR is the outermost among
first-order resonances -- we hence exclusively focus on this case.
The critical
time scale for capture into 2:1 MMR of equal-mass satellites is given
by \citep{ogi13}
\begin{equation}
    t_{\rm m}^{\rm crit} = 2.5\times 10^{4}\left(
        \frac{M_s}{M_{\rm Io}}\right)^{-4/3}
        \left(\frac{M_p}{M_{\rm J}}\right)^{4/3}T_{\rm K}\,,
    \label{tm_crit_io}
\end{equation}
where $M_{\rm Io}$ is the mass of Io and $T_{\rm K}$ is the orbital
period of the satellite.  Here, we only consider satellites with equal
masses because Galilean satellites have similar masses.

In Model 1, the first satellite is located at about $8.5R_{\rm J}$
after the termination of migration. A satellite in 2:1 resonance with
the satellite at $8.5R_{\rm J}$ has an orbit at approximately
$14R_{\rm J}$. The corresponding migration timescale is $t_{\rm m}(14R_{\rm
  J})\simeq1500\,$yr, which is longer than the critical time scale for
the first satellite, $t_{\rm m}^{\rm crit}(8.5R_{\rm J})=97\,$yr.  This
means that the second satellite is captured in the resonance.
Similarly, the third satellite can be captured in the 2:1 resonance of
the second satellite at $21R_{\rm J}$ because $t_{\rm m}(21R_{\rm J})\simeq
2600\,{\rm yr}> t_{\rm m}^{\rm crit}(14R_{\rm J})\simeq 200\,$yr.  In this
way, we successfully build-up a system in Laplace resonances.

\begin{figure}[h]
    \plotone{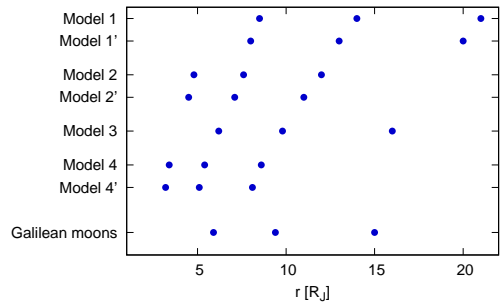}\\[-5pt]
        \caption{Comparison of the resonant orbits of the satellites
          obtained in our models with those of the Galilean moons.}
         \label{fig:system}
\end{figure}

The positions where the first satellite terminate for Models 2-4 are
$4.8R_{\rm J}$, $6.2R_{\rm J}$, and $3.4R_{\rm J}$, and the 2:1
resonance orbits of these are $7.6R_{\rm J}$, $9.8R_{\rm J}$, and
$5.4R_{\rm J}$, respectively.  The orbits in 2:1 resonance with the
second satellites are $12R_{\rm J}$, $16R_{\rm J}$, and $8.6R_{\rm
  J}$, respectively.  As summarized in Table \ref{tab:timescales},
the migration timescale of each of these orbits is larger than the critical
timescale for capture in the mean motion resonance. Thus, we can also
obtain systems in the Laplace resonance with Models~2-4, as well as
Model~1.  Similarly, we can form those systems in Models 1', 2' and
4'.  The comparison of orbits of Models 1-4, 1', 2' and 4' with the
Galilean moons are given in Figure~\ref{fig:system}.
Note that the orbits of the resonant three moons are located
on the same slope of the surface density profile in all models.
One can see that
Model~3 has a similar set of orbits with the inner three moons of the
Galilean system.

\begin{table}
    \centering
    \begin{tabular}{lccc}
      \hline
        & Migration timescale & & Critical timescale\\[2pt]
      \hline\hline\\[-6pt]
        Model 1 & $t_{\rm m}(14R_{\rm J})$ = 1500yr & $>$
                & $t_{\rm m}^{\rm crit}(8.5R_{\rm J})$ = 97yr \\
                & $t_{\rm m}(21R_{\rm J})$ = 2600yr & $>$
                & $t_{\rm m}^{\rm crit}(14R_{\rm J})$ = 200yr \\
                \hline
        Model 2 & $t_{\rm m}(7.6R_{\rm J})$ = 5000yr & $>$
                & $t_{\rm m}^{\rm crit}(4.8R_{\rm J})$ = 41yr \\
                & $t_{\rm m}(12R_{\rm J})$ = 7600yr & $>$
                & $t_{\rm m}^{\rm crit}(7.6R_{\rm J})$ = 82yr \\
                \hline
        Model 3 & $t_{\rm m}(9.8R_{\rm J})$ = 760yr & $>$
                & $t_{\rm m}^{\rm crit}(6.2R_{\rm J})$ = 60yr \\
                & $t_{\rm m}(16R_{\rm J})$ = 1300yr & $>$
                & $t_{\rm m}^{\rm crit}(9.8R_{\rm J})$ = 120yr \\
                \hline
        Model 4 & $t_{\rm m}(5.4R_{\rm J})$ = 2500yr & $>$
                & $t_{\rm m}^{\rm crit}(3.4R_{\rm J})$ = 24yr \\
                & $t_{\rm m}(8.6R_{\rm J})$ = 4000yr & $>$
                & $t_{\rm m}^{\rm crit}(5.4R_{\rm J})$ = 49yr \\
                \hline
    \end{tabular}
    \caption{Comparison between type-I satellite migration timescales,
      $t_{\rm m}$, and critical timescales, $t_{\rm m}^{\rm crit}$, for
      capture into MMR.}
    \label{tab:timescales}
    \smallskip
\end{table}

\section{Discussion}
\label{sec:discussion}

We summarize the orbits of satellites in our Model 3 along with
the mass and orbits of the inner three Galilean moons
that are in the Laplace resonance in Table \ref{tab:gal}.
\begin{table}[h]
    \centering
    \begin{tabular}{lccc}
      \hline
        \multicolumn{3}{c}{Galilean Moons}
        & \multicolumn{1}{c}{Model 3}\\[2pt]
        \hline\hline\\[-6pt]
            & Mass (10$^{25}$ g) & Orbit ($R_{\rm J}$)
            & Orbit ($R_{\rm J}$) \\
        \hline
        Io & 8.93 & 5.9 & 6.2 \\
        Europa & 4.80 & 9.4 & 9.8 \\
        Ganymede & 14.8 & 15.0 & 16 \\
        \hline
    \end{tabular}
    \caption{Summary of satellite properties}
    \label{tab:gal}
\end{table}
Systems in other models are more compact or spread-out compared with
the Galilean moons, but most importantly, the moons are in the 4:2:1
MMR.  Once they are in this resonance, the orbits are locked and the
moons migrate together as a system; the separations of the bodies
adjust accordingly, when the whole system moves radially during the
evolution of the CPD.

\begin{figure}[h]
    \plotone{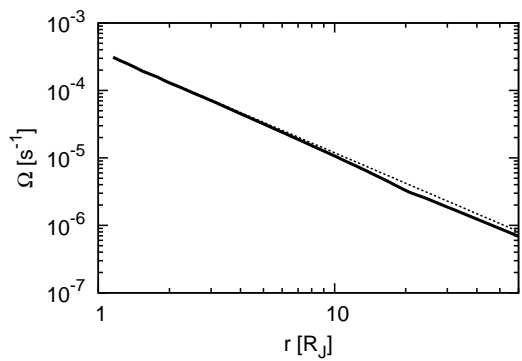}\\[-5pt]
        \caption{Angular velocity of the disk for Model 4.
        Keplerian frequency is also plotted in the dotted line.}
         \label{fig:omega}
\end{figure}

Figures \ref{fig:cpd-ep} and \ref{fig:cpd-ep-dep1} show that the disk
is quite hot at this stage. At such high disk temperatures, the
  radial pressure
gradient may lead to sub-Keplerian rotation velocities.
Actually, as Fig.~\ref{fig:omega} shows, the angular velocity is
smaller than the Keplerian value in the outer part of the disk.
The angular velocity is calculated as $\Omega = \Omega_{\rm K}(1-\eta)^{1/2}$,
where $\eta = -(r\Omega_{\rm K}^2\rho)^{-1}\partial P/\partial r$ \citep{tak02}.
We assumed Keplerian rotation profiles when we derived disk models.
However, since Equation (\ref{dSdt}) is only sensitive to the radial \emph{slope}
of the angular velocity, the assumption is expected to be acceptable.

Hot CPDs are suggested by \citet{kei14},
\citet{zhu15}, and \citet{szu16}, but it may be difficult to form icy
satellites in such an environment. Since the outer disk is cooler,
moons may gain icy materials simply by migrating in from larger radii.
Although Fig.~\ref{fig:b} shows $b$ is positive in $r \gtrsim 25R_{\rm
  J}$ for Model~4, for instance, bodies about ten times smaller than
Io can migrate all the way from the outer radii because $b$ for them
remains negative at all radii.  Another possibility is that ice-rich
planetesimals are captured when they enter into the CPD.
\citet{tan14} found that the orbits of sub-Io sized planetesimals
captured in a CPD are highly eccentric.  They also found that $10\,$m
or larger planetesimals can be efficiently captured in a CPD, thus
those bodies may grow into the size of present moons.

One problem is, however, whether the system can survive over the
long-term evolution of the CPD.  As mass infall decreases, the
temperature of the disk also decreases.  When the disk structure that
traps the innermost body disappears, the satellite system will start
to migrate toward the planet.  Moons can survive if the CPD is quickly
cleared before they are lost into the planet.  A rough estimate of the
viscous timescale of the disk is $r^2/\nu \sim 100\,$yr, which is
shorter than the migration timescale of the satellites.  However, the
actual timescale for the surface density to become small enough not
to affect satellite migration is most likely to be much longer than
this estimate. Clearly, this depends on how the disk dissipates and
many other unknown factors.  In order to obtain a better understanding
of how the infall terminates, we need to further study the evolution
of PPDs including both gap formation and gas dissipation.  In this
work, we adopted a mass infall rate derived from isothermal hydrodynamic
simulations, however, \citet{gre13} suggested that taking magnetic
field and radiative cooling into account leads to a different mass
infall rate, which opens a perspective for future work.

\citet{sas10} and \citet{ogi12} found that the existence of an inner
cavity in a CPD can prevent a moon system from being lost onto the
planet. Provided the planet rotates differentially (given that it
accretes material with non-negligible angular momentum this is not
unreasonable to assume) and maintains a convective or turbulent
sub-surface flow, it can be expected to harbor an efficient planetary
dynamo. In this case, such an inner cavity may form due to
magnetospheric truncation of the sub-disk by the planet's dipole
magnetic field. In the context of PPDs, not only photoevaporative but
also magnetically driven disk wind have been reported to contribute to
the formation of so-called ``transition disks'' \citep{suz10} with
reduced surface density at small radii.  To explore such currently
unknown effects in the context of embedded sub-disks, the
configuration of magnetic field at the very vicinity of the planet
must be studied.

\section{Summary}
\label{sec:summary}

We have modeled massive and comparatively hot CPDs by solving the time
evolution of surface density with mass infall from the parental PPD.
The mass infall flux was determined based on the high-resolution
numerical simulation of \citet{tan12}, where we have also considered
the reduction of the flux caused by the dissipation of the PPD at the
location of the sub-disk.  The temperature profile of the CPD is
derived by the balance of viscous heating and radiative cooling, as
well as the radial advection. Since the strength of viscosity is
uncertain in the absence of MRI, we employed a parameter to determine
the minimum value of the viscosity.  We considered the MRI when the
Elssasser number exceeds unity due to thermal ionization. We
furthermore monitored Toomre's Q parameter in order to consider
effective viscosity when the value becomes lower than about two. When
the evolution is governed by $\alpha_{\rm floor}$, the system settles
into a steady state.

In many previous studies, the critical temperature for the onset of
the MRI is assumed to be at about 1000$\,$K.  As shown in
Figure~\ref{fig:mri}, however, we found that this is not the case for
massive CPDs. This is because of the two reasons: (i) the ionization
degree needed to sustain the MRI in a CPD is higher than that in a
PPD, and (ii) thermal ionization is less effective in higher density
regions. In our models, MRI is turned on by thermal ionization only
around T$\sim2000-3000\,$K.

We found that opacity transitions change the radial dependence of the
temperature structure, and especially, a transition near 2000$\,$K
makes a bump in surface density distribution.  We estimated whether a
moon migrating toward the central planet can be trapped at such a
location. In the case of some of the parameter settings that are
referred to as Models 1-4, 1', 2', and 4', the surface-density and
temperature gradients were sufficiently steep to stop the migration of
a moon. Moreover, we have examined the migration timescales of the
second and third moons migrating inward and compared them to the
critical timescale to be captured in a 2:1 MMR with the inner moon.
In all of Models 1-4, 1', 2', and 4', we obtained systems in 4:2:1 mean
motion resonance that is known for inner three bodies of the Galilean
system.  The satellite system obtained in our disk models may or may
not survive until the dissipation of the CPD. In order to find out the
long term evolution of these systems, further studies on mass infall
from PPDs and on the origin of angular momentum transport in CPDs are
needed.

\acknowledgments

We thank the anonymous referee for a careful report.
We acknowledge Shigeo S. Kimura, Pablo Ben{\'i}tez-Llambay,
Shigeru Ida, Masahiro Ogihara, and Kazuhiro D. Kanagawa for fruitful
discussions and Edwin L. Turner for encouraging comments.
HK was supported by Grants-in-Aid for Scientific Research
(No. 26287101) from Ministry of Education, Culture, Sports,
Science and Technology (MEXT) of Japan and by Astrobiology Center
Project of the National Institute of Natural Science (NINS)
(Grant Number AB281018).
OG has received funding from the European Research Council (ERC)
under the European Union's Horizon 2020 research and innovation
programme (grant agreement No 638596).

\bibliographystyle{apj}

\end{document}